\documentstyle[aps,epsfig,floats,preprint,amsfonts]{revtex}
\begin{document}
\draft
\preprint{\vbox{{\hbox{\sf gr-qc/0409003}}}}
\title{Decay Rate and Low-Energy Near-Horizon Dynamics of\\
  Acoustic Black Holes}
\author
{Sung-Won Kim$^{a}$\footnote{email:sungwon@mm.ewha.ac.kr}, Won Tae
  Kim$^{b}$\footnote{email:wtkim@mail.sogang.ac.kr}, and John
  J. Oh$^{a,c}$\footnote{email:j4oh@sciborg.uwaterloo.ca}} 
\address{\small \it ${}^{a}$ Department of Science Education and Basic Science Research
  Institute,\\ Ewha Women's University, Seoul 120-750, Korea\\
         ${}^{b}$ Department of Physics and Basic Science Research
         Institute,\\ Sogang University, C.P.O. Box 1142, Seoul, Korea\\
         ${}^{c}$ Department of Physics, University of Waterloo, Waterloo, ON, N2L 3G1, Canada }
\maketitle
\begin{abstract}{\small 
We study the low-energy dynamics of an acoustic black hole near the sonic
horizon. For the experimental test of black hole
evaporation in the laboratory, the
decay rate (greybody factor) of the acoustic black hole (sonic
hole)  
can be calculated by the usual low-energy perturbation method. As
a consequence, we obtain the decay rate of the sonic horizon from the 
absorption and the reflection coefficients. Moreover, we show that the 
thermal emission from the sonic horizon is only proportional to a
control parameter which describes the velocity of the fluid.}
\end{abstract}     
\bigskip
\newpage
One of the most attractive aspects of general relativity is that it
predicts the existence of black holes. Classically, they can absorb
all sort of matter through the extremely strong gravitational field, and
even light cannot escape from them. For this
reason, they are black to all observers outside them. Their quantum
mechanical properties are quite different from their classical
ones. The surprising prediction that quantum black holes are no longer black
was announced in 1974 \cite{hawking}, thereafter the quantum mechanical
properties of them remain fascinating but problematic to theoretical and
experimental physicists. From the viewpoint of quantum field theory in
a curved spacetime background, they can emit thermal
radiation proportional to the surface gravity at their horizons.
This theoretical result, called ``Hawking radiation'', used to be asked
if it is possible to verify it from
experimental study within the framework of astrophysical phenomenology.

In 1981 Unruh proposed a possibility of an experimental test
for the black hole evaporation \cite{unruh}, which opened a new branch of
black hole physics \cite{nvv,cj,jv,jm,unruh2,us,visser,lsv,blv,fabs}. He
considered an irrotational inviscid fluid 
and sound wave propagating in the medium. According to
Ref. \cite{unruh}, the linearized small perturbation of wave equations
in the fluid leads to an equation of motion for a massless scalar field in the
geometrical background that is similar 
to the black hole metric. There exists a trapped region that the sound wave
cannot escape as the fluid velocity exceeds the sound velocity
(hypersonic flow), which is called a ``sonic horizon'' when the fluid
velocity is equals to that of the sound wave.
This hypersonic flow system reveals more aspects of the
acoustic black hole with the help of the analogy of black hole
thermodynamics. By quantizing the sound field (phonons), the
sonic hole (dumb hole) can emit sound waves with the thermal temperature given by
\begin{equation}
  \label{eq:hawk}
  T=\frac{\hbar}{2\pi} \left[\frac{\partial v^{r}}{\partial
  r}\right]_{\rm horizon},
\end{equation}
where $\hbar$ is a Planck constant and $v^{r}$ is a fluid velocity
along the radial direction. 

In this paper, we shall investigate the decay rate of sonic black holes from
the near-horizon low-energy dynamics, which is expected to give an
experimental suggestion of detecting the thermal radiation. The low-energy
perturbation analysis of phonons leads to an exact derivation of the 
absorption cross section and decay rate of the sonic horizon. As a
consequence, those results depend only upon a control parameter that
describes the velocity of the given fluid associated with the surface
gravity on the horizon, which agrees with Unruh's prediction of thermal radiation.

Let us briefly review the hypersonic flow and acoustic black holes
following Ref. \cite{unruh}. The irrotational inviscid fluid flow
is described by the following three equations,
\begin{eqnarray}
  \label{eq:eqnsfluid}
  & & \vec{\nabla} \times \vec{v} = 0, \nonumber \\
  & & \rho\left[ \frac{\partial}{\partial t} \vec{v} + (\vec{v}\cdot
  \vec{\nabla})\vec{v}\right] = - \vec{\nabla}p -
  \rho\vec{\nabla}\Phi, \nonumber \\
  & & \frac{\partial}{\partial t} \rho + \vec{\nabla}\cdot
  (\rho\vec{v}) = 0,
\end{eqnarray}
where $\vec{v}$ is a velocity of the flow, $p$ is a pressure that is a
function of the density $\rho$, and $\Phi$ is an external
potential. Note that the first equation describes an irrotational flow,
the second and the third ones are Eulerian and continuity equations,
respectively. These can be linearized in the vicinity of some mean flow solution
with $\xi=\xi_{0} + \bar{\xi}$ and $\psi = \psi_{0} + \bar{\psi}$
redefining the fields as $\vec{\nabla}G(\xi) = \vec{\nabla}p/\rho$,
$\rho = e^{\xi}$, and $\vec{v} = \vec{\nabla}\psi$, where the barred
parameters stand for the small fluctuations around the vacuum solutions
with a zero label. Combining these
linearized equations gives the massless scalar field equation
describing the sound wave in the curved spacetime background
\begin{equation}
  \label{eq:scalar}
  \frac{1}{\sqrt{-g}}\partial_{\mu}(\sqrt{-g}g^{\mu\nu}\partial_{\nu}
  \bar{\psi}) = 0,
\end{equation}
with the background metric,
\begin{equation}
  \label{eq:metric}
  g_{\mu\nu} =\frac{\rho_{0}}{c}\left(%
\begin{array}{cc}
  -c^2 + v_{0}^2 & -v_{0}^{i} \\
  -v_{0}^{j}& \delta_{ij} \\
\end{array}%
\right),
\end{equation}
where $i,j=1,2,3$ and $c$ is a local velocity of sound defined as $c^2
= G'(\xi)$. Following Ref. \cite{unruh}, we assume $c$ to be constant
for convenience.\footnote{This is assumed to be constant since the role of
  sound in the fluid is similar to that of light in gravity,
  aside from the easy treatment of the remaining calculations.}. Provided that the metric, (\ref{eq:metric}), is spherically symmetric and stationary,
we can  change the time coordinate $t$ into  $\tau = t+ \int dr (v_{0}^{r}(r)/(c^2
- {v_{0}^{r}}^2(r))$, which yields a diagonalized metric,
\begin{equation}
  \label{eq:diagmetric}
  (ds)^2 = \frac{\rho_{0}}{c} \left[ - (c^2 - {v_{0}^{r}}^2)d\tau^2 +
  \frac{c^2 dr^2}{c^2 - {v_{0}^{r}}^2} + r^2 d^2\Omega\right].
\end{equation}
 The coordinate transformation does not affect the remaining physical
 result although the metric is a coordinate-dependent quantity. We, therefore,
 can make it deform for convenience. Of course, if the metric had been
 transformed in the scalar field equation of Eq. (\ref{eq:scalar}),
 then the same diagonalized metric would have been obtained.
The metric (\ref{eq:diagmetric}) has a sonic horizon at $c^2 =
 {v_{0}^{r}}^2$, which coincides with the sonic fluid flow while the hypersonic flow ($c^2 <
{v_{0}^{r}}^2$) prevents the sound wave from 
propagating and the sound wave eventually
does not escape from the horizon.

If the flow smoothly exceeds the velocity of sound at the sonic horizon
$r=r_{sh}$, the velocity can be expanded as $v_{0}^{r}(r) = -c +
a(r-r_{sh}) + {\cal O}(r-r_{sh})^2$, which yields (up to the first
order of $r$)
\begin{equation}
  \label{eq:metcomp}
  c^2 - {v_{0}^{r}}^2 \approx 2ac(r-r_{sh}),
\end{equation}
where $a$ is a control parameter associated with the velocity of fluid
defined as $(\vec{\nabla} \cdot \vec{v})|_{r=r_{sh}}$. Note that the
parameter $a$ is related to the surface gravity up to some
proportional factor, which will be shown later.
The resulting metric, therefore, becomes
\begin{equation}
  \label{eq:metricbh}
  (ds)^2 = \frac{\rho_{0}}{c} \left[ - 2ac(r-r_{sh})d\tau^2 +
  \frac{cdr^2}{2a(r-r_{sh})} + r^2 d^2\Omega\right],
\end{equation}
which is similar to the near-horizon behavior of the Schwarzschild black
hole as discussed in Ref. \cite{unruh}. 

Our starting point is the field equation of the sound wave, Eq. (\ref{eq:scalar}),
under the near-horizon metric background of
Eq. (\ref{eq:metricbh}). Using the separation of variables, $\bar{\psi}(\tau,r,\theta,\varphi)
= u(r)\Theta(\theta)e^{i(m\varphi - \omega \tau)}$, the radial
equation of Eq. (\ref{eq:scalar}) is written as
\begin{equation}
  \label{eq:radialeq}
  (r-r_{sh})\partial_r^2 u(r) +\left[
  1+\frac{2}{r}(r-r_{sh})\right]\partial_{r} u(r) -
  \left(\frac{c\ell(\ell+1)}{2ar^2} -
  \frac{\omega^2}{4a^2(r-r_{sh})}\right) u(r)=0.
\end{equation}

What we wish to do is to approximate the wave equation in the
Schwarzschild geometry by another unphysical wave equation that has
the same behavior near the sonic horizon. Even though the unphysical
wave equation has the tremendous advantage that it is analytically
solvable, we will work within the first order of the approximation, 
because the near-horizon limit of the Schwarzschild metric gives the
simplified Eqs. (\ref{eq:metricbh}) and (\ref{eq:radialeq}). 

From the change of variables, $0 \le z= 1 - r_{sh}/r \le 1$, the
radial equation (\ref{eq:radialeq}) becomes
\begin{equation}
  \label{eq:zeq}
  z(1-z)\partial_{z}^2 u(z)+ \partial_{z}u(z) -
  \left[A - \frac{B}{z(1-z)}\right] u(z)=0,
\end{equation}
where $A=c\ell(\ell+1)/2ar_{sh}$ and $B=\omega^2/4a^2$. Now let us define $u(z) \equiv
z^{\alpha}(1-z)^{\beta}h(z)$ in order to remove singularities at $z=0$
and $z=1$, which leads to
\begin{equation}
  \label{eq:zeq2}
  z(1-z)\partial_{z}^2 h(z) + [ 1+ 2\alpha - 2(\alpha+\beta)z]
  \partial_{z} h(z) - [ A + (\alpha+\beta)(\alpha+\beta -1)] h(z) = 0,
\end{equation}
and the parameters $\alpha$ and $\beta$ are determined by $\alpha = \pm
i\sqrt{B}$ and $\beta = 1\pm\sqrt{1-B}$, respectively.
The solution of Eq. (\ref{eq:zeq2}) is simply given by the hypergeometric
function $F$ as
\begin{eqnarray}
  \label{eq:sol}
  & &u(z) = C_{1} z^{\alpha}(1-z)^{\beta}
  F\left(-\frac{1}{2}+\alpha+\beta-\frac{1}{2}\sqrt{1-4A},-\frac{1}{2}+\alpha+\beta+\frac{1}{2}\sqrt{1-4A};1+2\alpha;z\right)
  \nonumber \\    
  &+& C_{2} z^{-\alpha}(1-z)^{\beta}  F
  \left(-\frac{1}{2}-\alpha+\beta-\frac{1}{2}\sqrt{1-4A},-\frac{1}{2}-\alpha+\beta+\frac{1}{2}\sqrt{1-4A};1-2\alpha;z\right). 
\end{eqnarray}
Note that since this solution is symmetric for changing $\alpha$ into $
-\alpha$, we take the plus sign of $\alpha$ for convenience. In
$z\rightarrow 0$ limit, Eq. (\ref{eq:sol}) becomes
\begin{equation}
  \label{eq:near} 
  u_{\rm near}(r) \approx
  C_{\rm out}{\rm exp}\left[i\frac{\omega}{2a}\ln\left(1-\frac{r_{sh}}{r}\right)\right]
  + C_{\rm in}{\rm exp}\left[-i\frac{\omega}{2a}\ln\left(1-\frac{r_{sh}}{r}\right)\right],
\end{equation}
where $C_{\rm out}\equiv C_{1}$ and $C_{\rm in}\equiv C_{2}$. 
\begin{figure}[htbp]
    \begin{center}
    \leavevmode 
    \centerline{  
        \epsfig{figure=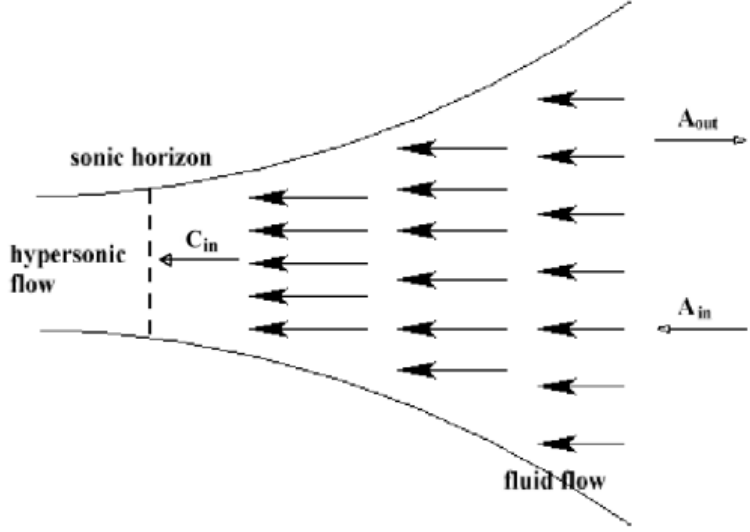, width=10cm, height=7.5cm}
               } 
    \caption{Configuration of the acoustic black holes and the boundary
  condition}
    \label{flows}
    \end{center}
\end{figure}
At this stage, the boundary condition can be imposed with an
appropriate physical situation, and here we take $C_{\rm out} =
0$. The configuration of the boundary condition that we are
considering is shown in the FIG. \ref{flows}. Using the $z\rightarrow 1-z$ transformation of hypergeometric
function \cite{as},
\begin{eqnarray}
  \label{eq:z1z}
  F(l,m;n;z) &=&
  \frac{\Gamma(n)\Gamma(n-l-m)}{\Gamma(n-l)\Gamma(n-m)}F(l,m;l+m-n+1;1-z)\nonumber \\
  &+& (1-z)^{n-l-m}\frac{\Gamma(n)\Gamma(l+m-n)}{\Gamma(l)\Gamma(m)} F(n-l;n-m;n-l-m+1;1-z),
\end{eqnarray}
the far-region solution of Eq. (\ref{eq:sol}) for $z\rightarrow 1$ is
given by
\begin{eqnarray}
  \label{eq:farsol}
  u_{\rm far}(r) &\approx& C_{\rm in} \left[
  \frac{\Gamma(1-2\alpha)\Gamma(2-2\beta)r_{sh}^{\beta}}{\Gamma\left(\frac{3}{2}-\alpha-\beta+\frac{1}{2}\sqrt{1-4A}\right)\Gamma\left(\frac{3}{2}-\alpha-\beta-\frac{1}{2}\sqrt{1-4A}\right)}r^{-\beta} 
  \right.\nonumber \\ 
  &+& \left.\frac{\Gamma(1-2\alpha)\Gamma(2\beta-2)r_{sh}^{2-\beta}}{\Gamma\left(-\frac{1}{2}-\alpha+\beta-\frac{1}{2}\sqrt{1-4A}\right)\Gamma\left(-\frac{1}{2}-\alpha+\beta+\frac{1}{2}\sqrt{1-4A}\right)}r^{\beta-2} \right].
\end{eqnarray} 
This far region solution is symmetric for changing $\beta$ into $2-\beta$,
which means that we can take the plus sign of $\beta$ for convenience. Since $\beta=1+\sqrt{4a^2 - \omega^2}/2a$, we consider
the real value case of $\beta$, which signifies that the control parameter
of the fluid velocity, $a$, is much greater than the energy of sound
wave, $4a^2\ge\omega^2$. This is nothing but a low-energy situation
that we are considering.

On the other hand, the asymptotic behavior of
Eq. (\ref{eq:radialeq}) produces the wave equation at boundary,
\begin{equation}
  \label{eq:beq} 
  r^2 \partial_{r}^2 u_{\rm B}(r) + 3 r\partial_{r} u_{\rm B}(r) +
  \frac{\omega^2}{4a^2}u_{\rm B}(r) = 0,
\end{equation}
which can be solved by
\begin{equation}
  \label{eq:bsol}
  u_{\rm B}(r) = A_{1} r^{-\beta} + A_{2} r^{\beta-2}.
\end{equation}

The coefficients, $C_{\rm in}$ and $A_{1}/A_{(2)}$, can be exactly
matched by comparing Eq. (\ref{eq:farsol}) with
Eq. (\ref{eq:bsol}). Note that $\beta = 2$ at low frequency region
since $4a^2 >> \omega^2$. The ``ingoing'' and ``outgoing'' waves (denoted by $A_{\rm
  in}$ and $A_{\rm out}$, respectively) 
of Eq. (\ref{eq:bsol}) can be decomposed by introducing the
redefinition of amplitudes,
\begin{eqnarray}
  \label{eq:redfamp}
  A_{1} &=& iL(A_{\rm in} - A_{\rm out}), \nonumber \\
  A_{2} &=& A_{\rm in} + A_{\rm out},
\end{eqnarray}
which leads to
\begin{equation}
  \label{eq:iobsol}
  u_{\rm B}(r) = u_{\rm B}^{\rm in}(r) + u_{\rm B}^{\rm out}(r)=
  A_{\rm in} \left(1+i\frac{L}{r^2}\right) + A_{\rm out} \left(1-i\frac{L}{r^2}\right),
\end{equation}
where $L$ is a positive dimensionless constant, which will be taken
to be independent of the energy $\omega$ \cite{bss,ko}. Note that $L$
can be chosen so that the absorption cross-section has the value of
the area of the black hole as $\omega \rightarrow 0$ \cite{bss,dgm} or it
has the usual value of the Hawking temperature \cite{ko}.
However, in this approach, we shall take into account a different requirement
of determining a positive dimensionless constant $L$. The fact that
${\cal A} + {\cal R} = 1$ can determine the dimensionless
constant $L$ in our configuration shown in FIG. \ref{flows}.

The absorption (${\cal A}$) and the reflection (${\cal R}$)
coefficients can be defined by the ratio of ``ingoing'' and
``outgoing'' fluxes, 
\begin{equation}
  \label{eq:absref}
  {\cal A} = \left|\frac{{\cal F}_{\rm near}^{\rm in}}{{\cal F}_{\rm
  B}^{\rm in}}\right|, ~~ {\cal R} = \left|\frac{{\cal F}_{\rm B}^{\rm out}}{{\cal F}_{\rm
  B}^{\rm in}}\right|.
\end{equation}
respectively.
It is easy to derive the flux from Eq. (\ref{eq:radialeq}) and it becomes
\begin{equation}
  \label{eq:flux}
  {\cal F} = \frac{2\pi}{i}r^2 g^{rr} (u^{*}(r)\partial_{r}u(r) - u(r)\partial_{r}u^{*}(r)).
\end{equation}
The reflection coefficient can be expressed in terms of the matched coefficients
in Eqs. (\ref{eq:farsol}) and (\ref{eq:bsol}) as
\begin{equation}
  \label{eq:ref}
  {\cal R} =  \left|\frac{{\cal F}_{\rm B}^{\rm out}}{{\cal F}_{\rm
  B}^{\rm in}}\right| = \left|\frac{A_{\rm out}}{A_{\rm
  in}}\right|^2 = \frac{L^2 |A_{2}|^2 +|A_{1}|^2 +iL(A_{2}^{*}A_{1} -
  A_{1}^{*}A_{2})}{L^2 |A_{2}|^2 +|A_{1}|^2 -iL(A_{2}^{*}A_{1} -
  A_{1}^{*}A_{2})}. 
\end{equation}
For $\ell = 0$ and $\omega \rightarrow 0$, the $|A_{2}|^2$ term in
Eq. (\ref{eq:ref}) is evaluated by
\begin{eqnarray}
  \label{eq:A2}
  |A_{2}|^2 &=& |C_{\rm in}|^2 \frac{r_{sh}^{4-2\beta}
   \Gamma(1-2\alpha)\Gamma(1+2\alpha)
   \Gamma^2(2\beta-2)}{\Gamma(\omega_{a} -
   \alpha)\Gamma(1+\omega_{a}-\alpha)\Gamma(\omega_{a}+\alpha)\Gamma(1+\omega_{a}+\alpha)}
   \nonumber \\
  &\approx& |C_{\rm in}|^2 r_{sh}^2,
\end{eqnarray}
by using $\Gamma(z)\Gamma(1-z)=\pi/\sin\pi z$. 
This approximation means that the Compton wavelength of the
scattered wave is much greater than the sonic black hole
size. Furthermore, the dominant contribution of $s$-mode is the case
of vanishing angular potential($\ell=0$) of the whole propagation
modes \cite{koy}. 
Similarly, the $|A_{1}|^2$ term can be calculated as $|A_{1}|^2 \approx
|C_{\rm in}|^2 r_{sh}^4$ in the same limit used in
Eq. (\ref{eq:A2}). Provided that this is compared with Eq. (\ref{eq:A2}), $|A_{1}|^2$
is negligible for the small sonic black hole horizon.

\begin{figure}[htbp]
    \begin{center}
    \leavevmode
    \centerline{
        \epsfig{figure=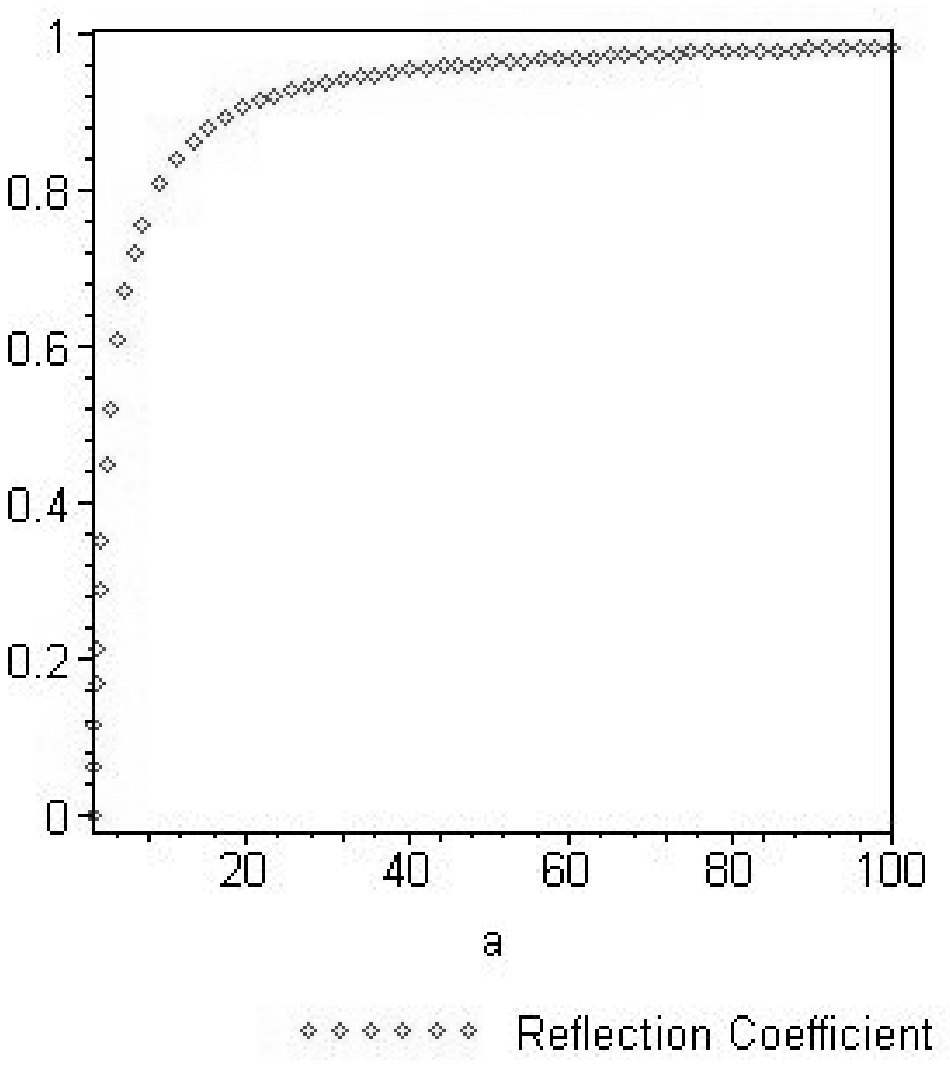, width=7cm, height=7cm}
        \epsfig{figure=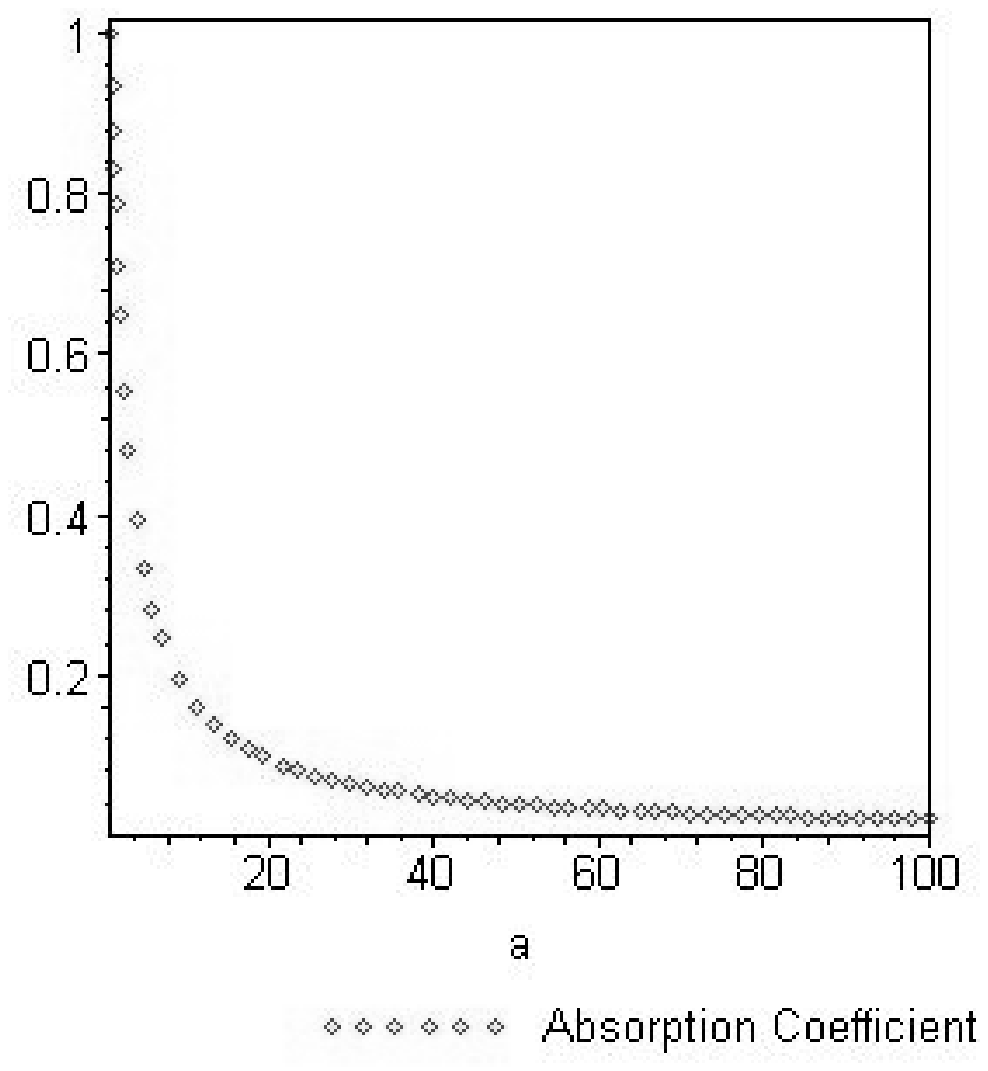, width=7cm, height=7cm}  
        } 
    \caption{Plots of the reflection and the absorption coefficients
    for $\omega = 1$. The valid range of $a$ is $a \ge 1$ for the
    fixed $\omega =1$.}
    \label{refnabs}
    \end{center}
\end{figure}

On the other hand, the non-trivial contribution of the reflection
coefficient arises from the term of $A_{2}^{*}A_{1} -
A_{1}^{*}A_{2}$. Using $\Gamma(z)\Gamma(1-z)=\pi/\sin\pi z$, we obtain
\begin{equation}
  \label{eq:cross}
  A_{2}^{*}A_{1} -A_{1}^{*}A_{2} \approx i\frac{r_{sh}^2\omega |C_{\rm in}|^2}{2a}.
\end{equation}
From Eqs. (\ref{eq:A2}) and (\ref{eq:cross}), the reflection
coefficient is simply expressed by
\begin{equation}
  \label{eq:reflec}
  {\cal R}^{\ell=0} \approx \frac{2aL-\omega}{2aL+\omega},
\end{equation}
which is expected to have a range with $0\le {\cal R} \le 1$ by definition. This
fact restricts the frequency of $\omega$ to $0\le \omega \le
2aL$. 
The absorption coefficient can be obtained by the reflection
coefficient since we have $1-{\cal R} = {\cal A}$, which produces
${\cal A}^{\ell=0} = 2\omega/(2aL+\omega)$. From the definition of absorption
coefficient in Eq. (\ref{eq:absref}), we also obtain ${\cal A}^{\ell=0} =
\omega/2L^2(2aL+\omega)$. Comparing these both relations determines the
positive dimensionless constant, $L=1/2$. As a consequence, the
reflection and the absorption coefficients are expanded as 
\begin{equation}
  \label{eq:absrefixL}
  {\cal R}^{\ell=0} = 1 - \frac{2\omega}{a} + {\cal
  O}(\omega^2),~~{\cal A}^{\ell=0} = \frac{2\omega}{a} + {\cal O}(\omega^2),
\end{equation}
and their behaviors are plotted in the FIG. \ref{refnabs}.
Note that the critical value of $a_{\rm c}=2\omega$ exists, which gives
``the total absorption'' (${\cal A}^{\ell=0} = 1$) and ``the no
reflection'' (${\cal R}^{\ell=0}=0$). This signifies that we have to
adjust the relation between the velocity and the energy of phonons, when we control
the velocity of the fluid in the experimental system.

On the other hand, a relation between the vacuum expectation value
of the number operator and the reflection coefficient can be given in the form \cite{gl,ko},
\begin{equation}
  \label{eq:reltr}
  <0|N|0>= \left(\left|\frac{A_{\rm in}}{A_{\rm out}}\right|^2 -1\right)^{-1}=\frac{{\cal R}}{1-{\cal R}}.
\end{equation}
Finally, the Hawking temperature is defined by
\begin{equation}
  \label{eq:hawkingtem}
  <0|N|0> = \frac{1}{{\rm exp}(\omega/T_{H})-1},
\end{equation}
and combining this with Eq. (\ref{eq:absrefixL}) leads to the Hawking
temperature as 
\begin{equation}
  \label{eq:hawkingtem}
  T_{H}|_{\omega\rightarrow 0} \approx \frac{a}{2}.
\end{equation}
This result coincides with the well-known result of the Hawking temperature in that
it depends only upon the flow divergence($a$) passing through the
sonic horizon surface at a certain time, which is closely related to surface
gravity on the horizon. 
The absorption cross section for the dominant s-wave mode is
given by $\sigma_{\rm abs}^{\ell =0} =
{\cal A}/\omega = 2/(a+\omega)$ with the help of
Eq. (\ref{eq:absrefixL}), and the decay rate of the sonic horizon can be
easily calculated by
\begin{equation}
  \label{eq:decayr}
  \Gamma^{\ell = 0}_{\rm decay}  = \left.\frac{\sigma_{\rm
  abs}^{\ell=0}}{e^{\omega/T_{H}} - 1}\right|_{\omega \rightarrow 0} \approx \frac{4}{\omega}.
\end{equation}

The system we are considering is only for the low-energy frequency of
the phonon field. However, at high frequencies, there exist a kind of backscattering from the
geometry of the acoustic black hole. More precisely, this is
equivalent to the case of $\omega^2 \ge 4 a^2$ in our approach, which
yields an imaginary value of $\beta$. A straightforward calculation
following a similar procedure at high frequencies produces the exact
results for the reflection and the absorption coefficients,
\begin{eqnarray}
  \label{eq:highf}
  & &{\cal R}_{\omega>>}=\frac{\cosh \pi\omega_{a}^{-} + \cos
  {j}_{\ell}\pi}{\cosh \pi\omega_{a}^{+} + \cos {j}_{\ell}\pi},
  \nonumber \\
  & &{\cal A}_{\omega>>} = \frac{\cosh \pi\omega_{a}^{+} - \cosh \pi
  \omega_{a}^{-}}{\cosh \pi\omega_{a}^{+} + \cos {j}_{\ell}\pi}, 
\end{eqnarray}
respectively. Note that $\omega_{a}^{\pm} = \omega/a \pm
\sqrt{\omega^2-4a^2}/a$ and $j_{\ell} =
\sqrt{1-\frac{2c}{ar_{sh}}\ell(\ell+1)}$ and it is easy to show that
${\cal R}_{\omega>>} +{\cal A}_{\omega>>} = 1$. Since
$\omega_{a}^{+} \approx 2\omega/a$ and $\omega_{a}^{-} \approx 0$ at
high frequency region, the reflection (${\cal R}$) and the
absorption coefficients (${\cal A}$) are
\begin{eqnarray}
  \label{eq:highbehav}
  & & {\cal R}_{\omega >>} = \left.\frac{1+\cos
  j_{\ell}\pi}{\cosh{2\pi\omega}{a} + \cos j_{\ell}\pi}\right|_{\omega>>}
  \rightarrow 0, \nonumber \\
  & & {\cal A}_{\omega >>} = \left.\frac{\cosh{2\pi\omega}{a} - 1}{\cosh{2\pi\omega}{a} + \cos j_{\ell}\pi} \right|_{\omega>>}
  \rightarrow 1.
\end{eqnarray}
Note that this behavior is nothing but a scattering problem in
the flat spacetime background. In the tortoise
coordinate system, $r^{*} = \int dr/\sigma(r)$, one may read off the
effective potential given by
$V_{\rm eff} (r) = \omega^2 - V_{\ell} (r)$ from the Wheeler-Regge equation, where 
\begin{equation}
  \label{eq:effepoten}
  V_{\ell}(r) = \sigma(r) \left[\frac{\partial_{r}\sigma(r)}{r} + \frac{\ell(\ell+1)}{r^2}\right]
\end{equation}
and $\sigma(r)$ is a metric function in the metric, $(ds)^2 = -\sigma(r)dt^2 +
 \sigma^{-1}(r)dr^2 +d^2\Omega$.
It is easily shown that the effective potential of the wave equation becomes trivial
as $V_{\rm eff}(r)= \omega^2$ at high frequencies. Therefore, the Wheeler-Regge equation
describes the free field equation with the constant effective potential of flat spacetime. Furthermore, the decay rate at high frequencies from
Eq. (\ref{eq:decayr}) vanishes as $\omega$ goes to infinity, which
is compatible with the statements at high frequencies as discussed before.
\acknowledgments{J.J.O. would like to thank Edwin J. Son, Jaehoon Kim,
  and Hyunjoo Lee for exciting and valuable discussions and he is
  grateful to Yoonbai Kim, Hyeonjoon Shin, and Gungwon Kang for a
  warm hospitality while he was staying at SKKU and KIAS as a visiting
  scholar. He also wishes to thank Russell J. Smith for helpful
  comments on this paper. This work was supported by the Basic Science
  Research Program of the Korean Science \& Engineering Foundation
  Grant. At the final stage of this work, J.J.O. was supported by the
  Post-doctoral Fellowship Program of Korea Science \& Engineering
  Foundation (KOSEF). 
}


\end{document}